\documentclass[12pt,twoside,fleqn]{report}
\usepackage{epsfig}
\usepackage{graphics}
\usepackage{graphicx}
\usepackage{subfigure}
\usepackage{exscale}
\usepackage{latexsym}
\usepackage{makeidx}
\usepackage{amsfonts}
\usepackage{amsmath}
\usepackage{xtab}
\usepackage[section]{placeins}
\begin{document}

\newcounter{univ_counter}
\setcounter{univ_counter} {0}
\addtocounter{univ_counter} {1} 
\edef\MSU{$^{\arabic{univ_counter}}$ } \addtocounter{univ_counter} {1} 
\edef\INFNGE{$^{\arabic{univ_counter}}$ } \addtocounter{univ_counter} {1} 
\edef\ROMATRE{$^{\arabic{univ_counter}}$ } \addtocounter{univ_counter} {1} 
\edef\RPI{$^{\arabic{univ_counter}}$ } \addtocounter{univ_counter} {1} 
\edef\SACLAY{$^{\arabic{univ_counter}}$ } \addtocounter{univ_counter} {1} 
\edef\WM{$^{\arabic{univ_counter}}$ } \addtocounter{univ_counter} {1} 
\edef\UMASS{$^{\arabic{univ_counter}}$ } \addtocounter{univ_counter} {1} 
\edef\JLAB{$^{\arabic{univ_counter}}$ } \addtocounter{univ_counter} {1} 
\edef\INFNFR{$^{\arabic{univ_counter}}$ } \addtocounter{univ_counter} {1} 
\edef\YEREVAN{$^{\arabic{univ_counter}}$ } \addtocounter{univ_counter} {1} 
\edef\ASU{$^{\arabic{univ_counter}}$ } \addtocounter{univ_counter} {1}
\edef\ORSAY{$^{\arabic{univ_counter}}$ } \addtocounter{univ_counter} {1} 
\edef\EDINBURGH{$^{\arabic{univ_counter}}$ } \addtocounter{univ_counter} {1} 
\edef\GWU{$^{\arabic{univ_counter}}$ } \addtocounter{univ_counter} {1} 
\edef\UNH{$^{\arabic{univ_counter}}$ } \addtocounter{univ_counter} {1} 
\edef\OHIOU{$^{\arabic{univ_counter}}$ } \addtocounter{univ_counter} {1} 
\edef\CMU{$^{\arabic{univ_counter}}$ } \addtocounter{univ_counter} {1} 
\edef\CUA{$^{\arabic{univ_counter}}$ } \addtocounter{univ_counter} {1} 
\edef\SCAROLINA{$^{\arabic{univ_counter}}$ } \addtocounter{univ_counter} {1} 
\edef\ODU{$^{\arabic{univ_counter}}$ } \addtocounter{univ_counter} {1} 
\edef\UTEP{$^{\arabic{univ_counter}}$ } \addtocounter{univ_counter} {1} 
\edef\VIRGINIA{$^{\arabic{univ_counter}}$ } \addtocounter{univ_counter} {1} 
\edef\PITT{$^{\arabic{univ_counter}}$ } \addtocounter{univ_counter} {1} 
\edef\FSU{$^{\arabic{univ_counter}}$ } \addtocounter{univ_counter} {1} 
\edef\CNU{$^{\arabic{univ_counter}}$ } \addtocounter{univ_counter} {1} 
\edef\VT{$^{\arabic{univ_counter}}$ } \addtocounter{univ_counter} {1} 
\edef\DUKE{$^{\arabic{univ_counter}}$ } \addtocounter{univ_counter} {1} 
\edef\UCONN{$^{\arabic{univ_counter}}$ } \addtocounter{univ_counter} {1} 
\edef\MIT{$^{\arabic{univ_counter}}$ } \addtocounter{univ_counter} {1} 
\edef\URICH{$^{\arabic{univ_counter}}$ } \addtocounter{univ_counter} {1} 
\edef\JMU{$^{\arabic{univ_counter}}$ } \addtocounter{univ_counter} {1} 
\edef\GLASGOW{$^{\arabic{univ_counter}}$ } \addtocounter{univ_counter} {1} 
\edef\NSU{$^{\arabic{univ_counter}}$ } \addtocounter{univ_counter} {1} 
\edef\KYUNGPOOK{$^{\arabic{univ_counter}}$ } \addtocounter{univ_counter} {1} 
\edef\ITEP{$^{\arabic{univ_counter}}$ } \addtocounter{univ_counter} {1} 
\edef\RICE{$^{\arabic{univ_counter}}$ } \addtocounter{univ_counter} {1} 
\edef\FIU{$^{\arabic{univ_counter}}$ } \addtocounter{univ_counter} {1} 
\edef\UCLA{$^{\arabic{univ_counter}}$ } \addtocounter{univ_counter} {1} 

\begin{titlepage}


\begin{flushright}
CLAS-NOTE 2003-001
\end{flushright}

\vspace{0.3 in}

\begin{center}
{\LARGE\bf The proton structure function $F_2$ with CLAS}

\vspace{0.3 in}

{\small
M.~Osipenko,\MSU$^,$\INFNGE\
G.~Ricco,\INFNGE\
M.~Taiuti,\INFNGE\
M.~Ripani,\INFNGE\
S.~Simula,\ROMATRE\ 
G.~Adams,\RPI\
E.~Anciant,\SACLAY\
M.~Anghinolfi,\INFNGE\
B.~Asavapibhop,\UMASS\
G.~Audit,\SACLAY
% ...
\
\\{\Large\bf (The CLAS Collaboration)}}

\vspace{0.5 in}

{\MSU \small Moscow State University, 119992 Moscow, Russia} \\
{\INFNGE \small INFN, Sezione di Genova, and Dipartimento di Fisica dell'Universit\`a, 16146 Genova, Italy} \\
{\ROMATRE \small Universit\`a di ROMA III, 00146 Roma, Italy} \\
{\RPI \small Rensselaer Polytechnic Institute, Troy, New York 12180} \\
{\SACLAY \small CEA-Saclay, Service de Physique Nucl\'eaire, F91191 Gif-sur-Yvette, Cedex, France} \\
{\WM \small College of William and Mary, Williamsburg, Virginia 23187} \\
{\UMASS \small University of Massachusetts, Amherst, Massachusetts 01003} \\
{\JLAB \small Thomas Jefferson National Accelerator Facility, Newport News, Virginia 23606} \\
{\INFNFR \small INFN, Laboratori Nazionali di Frascati, PO 13, 00044 Frascati, Italy} \\
{\YEREVAN \small Yerevan Physics Institute, 375036 Yerevan, Armenia} \\
{\ASU \small Arizona State University, Tempe, Arizona 85287} \\
{\ORSAY \small Institut de Physique Nucleaire ORSAY, IN2P3 BP 1, 91406 Orsay, France} \\
{\EDINBURGH \small Edinburgh University, Edinburgh EH9 3JZ, United Kingdom} \\
{\GWU \small The George Washington University, Washington, DC 20052} \\
{\UNH \small University of New Hampshire, Durham, New Hampshire 03824} \\
{\OHIOU \small Ohio University, Athens, Ohio 45701} \\
{\CMU \small Carnegie Mellon University, Pittsburgh, Pennsylvania 15213} \\
{\CUA \small Catholic University of America, Washington, D.C. 20064} \\
{\SCAROLINA \small University of South Carolina, Columbia, South Carolina 29208} \\
{\ODU \small Old Dominion University, Norfolk, Virginia 23529} \\
{\UTEP \small University of Texas at El Paso, El Paso, Texas 79968} \\
{\VIRGINIA \small University of Virginia, Charlottesville, Virginia 22901} \\
{\PITT \small University of Pittsburgh, Pittsburgh, Pennsylvania 15260} \\
{\FSU \small Florida State University, Tallahassee, Florida 32306} \\
{\CNU \small Christopher Newport University, Newport News, Virginia 23606} \\
{\VT \small Virginia Polytechnic Institute and State University, Blacksburg, Virginia   24061} \\
{\DUKE \small Duke University, Durham, North Carolina 27708} \\
{\UCONN \small University of Connecticut, Storrs, Connecticut 06269} \\
{\MIT \small Massachusetts Institute of Technology, Cambridge, Massachusetts  02139} \\
{\URICH \small University of Richmond, Richmond, Virginia 23173} \\
{\JMU \small James Madison University, Harrisonburg, Virginia 22807} \\
{\GLASGOW \small University of Glasgow, Glasgow G12 8QQ, United Kingdom} \\
{\NSU \small Norfolk State University, Norfolk, Virginia 23504} \\
{\KYUNGPOOK \small Kyungpook National University, Taegu 702-701, South Korea} \\
{\ITEP \small Institute of Theoretical and Experimental Physics, Moscow, 117259, Russia} \\
{\RICE \small Rice University, Houston, Texas 77005} \\
{\FIU \small Florida International University, Miami, Florida 33199} \\
{\UCLA \small University of California at Los Angeles, Los Angeles, California  90095} \\

\end{center}
\end{titlepage}

\date{\today}

The inclusive, inelastic $ep$ scattering cross section has
been measured with the CLAS~\cite{CLASNIM} detector
in Hall B of the
Thomas Jefferson National Accelerator Facility (TJNAF).

With these data
and other world data, we have extracted the Nachtmann
moments~\cite{Nachtmann} of $F_2$ and higher twist
parameters for $0.05 < Q^2 < 100$~GeV$^2$/c$^2$.  These results are
published in Ref.~\cite{mymoments}.  The purpose of this CLAS-Note is to
tabulate the CLAS $F_2$ data.
A description of the analysis is presented in the paper~\cite{mymoments}.

\begin{figure}[!h]
\begin{center}
\epsfig{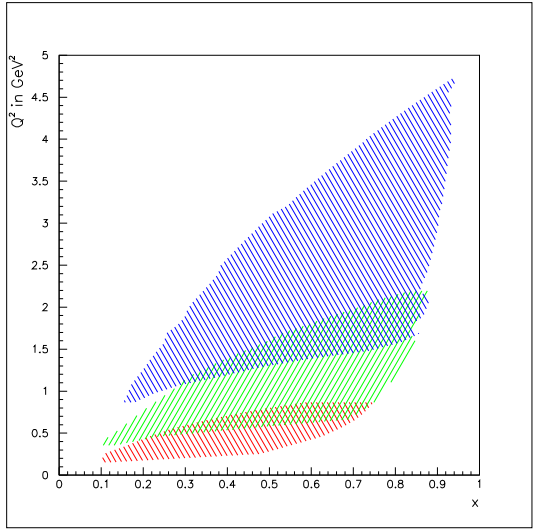}
\caption{Kinematic regions covered by different data sets:
red area - $E=1.5$~GeV;
green area - $E=2.5$~GeV;
blue area - $E=4$~GeV.}
\label{fig:d_ad1}
\end{center}
\end{figure}

We used reconstructed data taken with CLAS on a hydrogen target
collected during February and March 1999 electron beam running period.
The present analysis is based on the data sets listed in Table~\ref{table:d_ad1}.
The kinematic range of the data is shown in Fig.~\ref{fig:d_ad1}.
\begin{table}[!h]
\begin{center}
\caption{Total number of electron events.}
\label{table:d_ad1}
\vspace{2mm}
\begin{tabular}{|c|c|c|} \hline
\multicolumn{2}{|c|}{Data set} & Total number of electron events \\ \hline
Energy [GeV] & Torus current [A] & x$10^6$ \\ \hline
1.5 & 1500 & 131 \\ \hline
1.5 & 2250 &  51 \\ \hline
2.5 (unpolarized beam) & 1500 &  17 \\ \hline
2.5 & 1500  &  125 \\ \hline
2.5 (unpolarized beam) & 2250 &  29 \\ \hline
2.5 & 2250 &  100 \\ \hline
4.0 & 2250 &  80 \\ \hline
4.2 & 2250 &  92 \\ \hline
4.4 & 2250 &  58 \\ \hline
\end{tabular}
\end{center}
\end{table}

\begin{figure}[!h]
\centering
\subfigure[$Q^2=1.1$~GeV$^2$/c$^2$]{
\includegraphics[width=8cm]{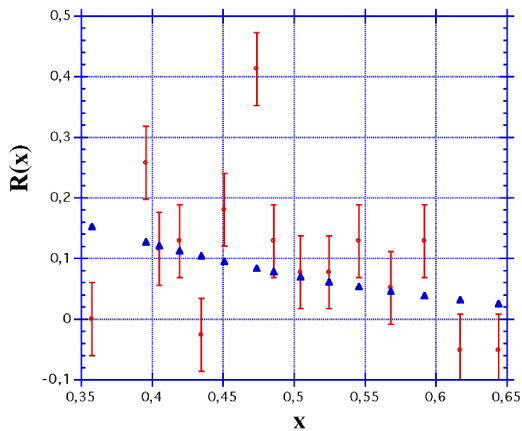}}
\subfigure[$Q^2=2$~GeV$^2$/c$^2$]{
\includegraphics[width=8cm]{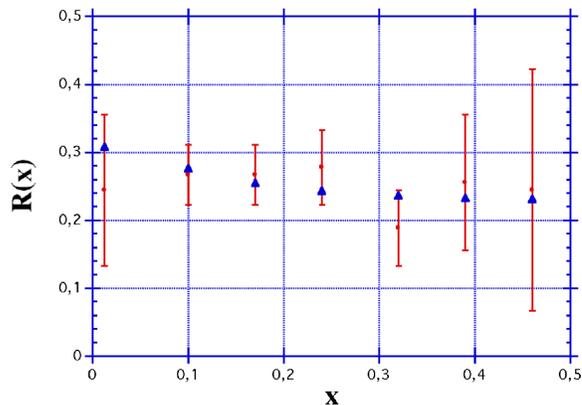}}
\caption{The fit of ratio $R$ at small $Q^2$:
red points - experimental data~\cite{BurkertPP,Ratio1,Ratio2};
blue points - fit developed in~\cite{Ricco1}.}
\label{fig:d_sf1}
\end{figure}

The structure function $F_2(x,Q^2)$ was extracted from the
inelastic cross section as follows:
\begin{equation}
F_2(x,Q^2)=\frac{1}{\sigma_{Mott}}\frac{d^2\sigma}
{dx dQ^2}J
\frac{\nu}{1+\frac{1-\epsilon}{\epsilon}\frac{1}{1+R}}~,
\label{eq:d_sf1}
\end{equation}
\noindent where $\sigma_{Mott}=\frac{\alpha^2 \cos{\frac{\theta}{2}}}{4E^2\sin^4{\frac{\theta}{2}}}$, $J$ is the Jacobian given by
\begin{equation}
J=\frac{x(s-M^2)}{2\pi M\nu}E^{\prime}~,
\end{equation}
\noindent
the squared invariant mass of the initial electron-proton system $s=M^2+2EM$ and
$\epsilon$ is the polarization parameter defined as
\begin{equation}
\epsilon \equiv \Biggl(
1+2\frac{\nu^2+Q^2}{Q^2} \tan^2{{\theta \over 2}}
\Biggr )^{-1}~.
\label{eq:d_sf2}
\end{equation}
\noindent The evaluation of $F_2(x,Q^2)$ was performed using the fit of the ratio
$R(x,Q^2) \equiv \sigma_L/\sigma_T$ developed in Ref.~\cite{Ricco1}.
The function $R(x,Q^2)$ was described as follows:
\begin{equation}
R(x,Q^2)= \left\{
\begin{array}{ll}
\frac{(1-x)^3}{(1-x_{th})^3}\bigl(0.041
\frac{\xi_{th}}{\log{\frac{Q^2}{0.04}}}+
\frac{0.592}{Q^2}-\frac{0.331}{(0.09+Q^4)}\bigr) & W < 2.5 ~~,\\
0.041\frac{\xi}{\log{\frac{Q^2}{0.04}}}+
\frac{0.592}{Q^2}-\frac{0.331}{(0.09+Q^4)} & W > 2.5 ~~,
\end{array}
\right.
\label{eq:d_sf3}
\end{equation}
\noindent and consists of two parts:
the fit for the DIS region ($W > 2.5$ GeV)~\cite{Whitlow_R,bar} and
a parameterization over the resonance region ($W < 2.5$ GeV),
where the data are scarce at small $Q^2$~\cite{BurkertPP,Ratio1,Ratio2}.
New measurement of $R$ in the resonance region undergoing in Hall~C TJNAF
will eliminate this lack of knowledge of the longitudinal cross section.
The systematic error on this parametrization was estimated according to Ref.~\cite{Ricco1}
as follows:
\begin{equation}
\delta_R= \left\{
\begin{array}{ll}
0.08 & W < 2.5 ~,\\ \\
\frac{0.006\xi}{\zeta}+
\frac{0.01}{Q^2}-\frac{0.01}{(0.09+Q^4)} & W > 2.5 ~~,
\end{array}
\right.
\label{eq:d_sf4}
\end{equation}
\noindent
where
\begin{eqnarray}
&& \zeta = \log{\frac{Q^2}{0.04}} ~~, \\ \nonumber
&& \xi = 1+12\frac{Q^2}{1+Q^2}
\frac{0.015625}{0.015625+x^2} ~~, \\ \nonumber
&& \xi_{th} = \xi(W=2.5) \mbox{   and   } x_{th} = x(W=2.5) ~~. \nonumber
\label{eq:d_sf5}
\end{eqnarray}
\noindent All dimensional variables are given in GeV. An example of
this fit is shown in Figure~\ref{fig:d_sf1} for two values of $Q^2$."

\begin{figure}[!h]
\centering
\includegraphics[width=12cm]{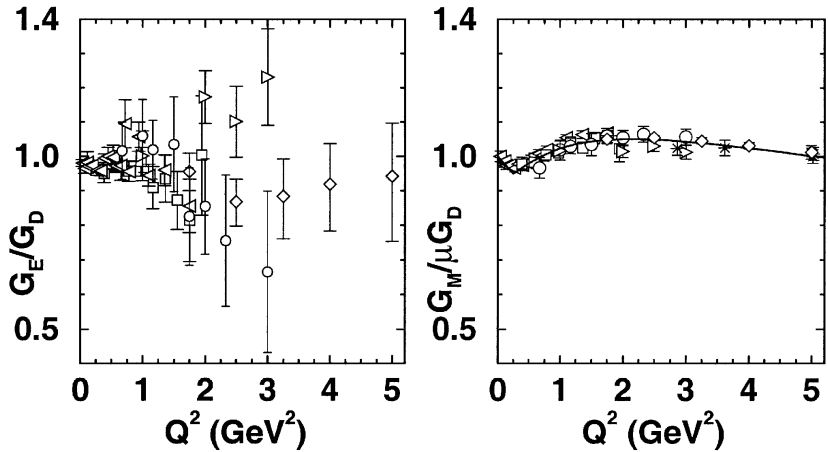}
\caption{World data for $G_E/G_D$ and $G_M/G_D$. The solid curve is a fit to
the $G_M$ data~\cite{Bosted}.}\label{fig:a1}
\end{figure}

\begin{figure}[!h]
\centering
\includegraphics[width=7cm]{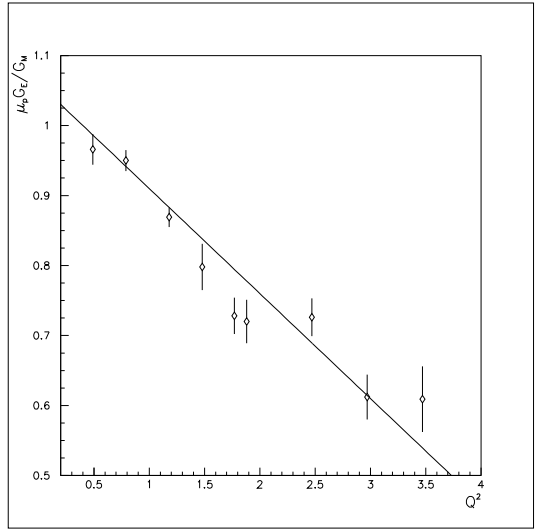}
\caption{Our new fit of the ratio $\frac{G_E}{G_M}$ shown as the solid
line with world data.}\label{fig:b1}
\end{figure}

Since the precision of the proton form factors is important for the present
analysis (they enter into the radiative correction procedure and into the
systematic
error estimates), recent high precision TJNAF Hall~A data on the ratio
$\frac{G_E}{G_M}$~\cite{hall-a}
forced us to modify the existing parametrization~\cite{Bosted}.
The magnetic form factor $G_M$ is very well established (see
Fig.~\ref{fig:a1}), whereas $G_E$ is not.
Therefore, the electric form factor $G_E$  from the
parametrization~\cite{Bosted} was adjusted to accommodate the new data.
This was accomplished by making a polynomial fit to the data
as follow:
\begin{equation}\label{eq:a1}
\frac{G_E}{G_M}(Q^2)=
\frac{G_{E_{Bosted}}(Q^2)(1-(A+BQ^2))}{G_{M_{Bosted}}(Q^2)} ~~,
\end{equation}
\noindent where $A$ and $B$ are free parameters. The resulting function
$G_E(Q^2)$ is shown in Fig.~\ref{fig:b1}. The corresponding
parameter values were found to be $A=-0.05$ and $B=0.625$.

This simple parameterization does not eliminate the inconsistency
of the polarization transfer method results from Ref.~\cite{hall-a} with other world data
obtained by means of the Rosenbluth separation,
but but it does combine the most precise data
from both methods. A possible small discrepancy with the
elastic electron scattering cross section in particular kinematic regions
does not affect the obtained structure function values.

\noindent In Table~\ref{tab:f2_clas} we tabulate $F_2(x,Q^2)$ extracted
from the CLAS data.

\setlength\textheight{750pt}
\setlength\voffset{-0.5in}
\setlength\topmargin{1pt}
\setlength\headheight{1pt}
\setlength\headsep{1pt}
\setlength\footskip{0pt}
\setlength\marginparwidth{1pt}
\renewcommand{\topfraction}{0.99}
\renewcommand{\bottomfraction}{0.99}


\begin{center}
\twocolumn
\topcaption{$F_2(x,Q^2)$ with statistical and systematic errors
extracted from CLAS data.}\label{tab:f2_clas}
\scriptsize
\tablehead{\hline}
\tabletail{\hline}
\begin{xtabular}{|c|c|c|c|c|} \hline
$Q^2$     & $x$ & $F_2$ & $\delta_{stat}$ & $\delta_{sys}$ \\ \hline
[GeV$^2$] &     &       &                 &                \\ \hline
0.225 & 0.1315 & 3.009E-01 & 6.194E-04 & 1.573E-02 \\ \cline{2-5}
 & 0.1405 & 3.228E-01 & 7.125E-04 & 1.276E-02 \\ \cline{2-5}
 & 0.1495 & 2.487E-01 & 6.064E-04 & 8.952E-03 \\ \cline{2-5}
 & 0.1585 & 2.053E-01 & 5.748E-04 & 8.110E-03 \\ \cline{2-5}
 & 0.1675 & 1.876E-01 & 6.007E-04 & 7.170E-03 \\ \cline{2-5}
 & 0.1765 & 1.794E-01 & 6.436E-04 & 5.482E-03 \\
\hline
\end{xtabular}
\end{center}

\end{document}